\documentclass[11pt,twoside]{article}
\usepackage{asp2010}

\resetcounters

\begin{document}

\title{Testing the binary hypothesis for the formation and shaping of planetary nebulae}

\author{D. Douchin$^{1,2}$, O. De Marco$^1$, G.~H. Jacoby$^3$, T.~C. Hillwig$^4$, D.~J. Frew$^1$, I. Bojicic$^1$, G. Jasniewicz$^2$ and Q.~A. Parker$^1$}
\affil{$^1$Astronomy, Astrophysics and Astrophotonics Research Center and Department of Physics and Astronomy, Macquarie University, Sydney, NSW 2122, Australia}
\affil{$^2$Laboratoire Univers et Particules de Montpellier (LUPM), UMR 5299 - CC72, Universit\'e Montpellier 2, Place Eug\`ene Bataillon, 34 095 Montpellier Cedex 5, France}
\affil{$^3$GMTO Corp and the Carnegie Observatories, Pasadena, CA 91101, USA}
\affil{$^4$Dept. of Physics and Astronomy, Valparaiso University, Valparaiso, IN 46383, USA}

\begin{abstract}
There is no quantitative theory to explain why a high 80\% of all planetary nebulae are non-spherical. The \emph{Binary Hypothesis} states that a companion to the progenitor of a central star of  planetary nebula is required to shape the nebula and even for a planetary nebula to be formed at all. A way to test this hypothesis is to estimate the binary fraction of central stars of planetary nebulae and to compare it with that of the main sequence population. Preliminary results from photometric variability and the infrared  excess techniques indicate that the  binary fraction of central stars of planetary nebulae is higher than that of the main sequence, implying that PNe could preferentially form via a binary channel. This article briefly reviews these results and current studies aiming to refine the binary fraction.
\end{abstract}



\section{Introduction}

Although it is a short phase of stellar evolution, planetary nebulae (PNe) are objects of interest, both because  they can be used to probe the physics of stars at a precise moment in their evolution, and because they are WD progenitors. In spite of their key role in Sun-like stars evolution, the mechanisms of their formation and shaping are uncertain to such an extent that we even doubt that they are really the product of all single Sun-like stars. Indeed, more than 80\% {\citep{ParkerEtAl-2006,JacobyEtAl-2010}} of PNe are non-spherical, showing structures such as lobes  and jets that give an axisymmetric, point-symmetric or asymmetric shape to the nebula. The hypothesis traditionally used to account for these shapes has been the action of global magnetic fields during the super wind phase of an AGB star, diverting the gas being ejected to the equatorial plane. However, this hypothesis has been contested by \citet{Soker-2006} and \citet{NordhausBlackmanAndFrank-2007}, who argued that the magnetic field could not be sustained for long enough on a whole-star scale due to the coupling between the magnetic field and the star rotation ; the field slows the rotation down and quenches itself. Another  hypothesis to account for non-spherical shapes of PNe is the presence of a companion (e.g. \citet{Soker-1997}). The hypothesis according to which a companion is required to shape an non-spherical PN has been dubbed the \emph{Binary Hypothesis} \citep{DeMarco-2009}. To test it, a necessary step is to estimate the  binary fraction of central stars of planetary nebulae (CSPNe). If the observed binary fraction of the CSPN population is superior to that of the putative parent population (the main sequence (MS) stars with mass $\simeq$~1-8~M$_{\odot}$, \citealt{MoeAndDeMarco-2006}), this indicates that PNe are preferentially  a binary phenomenon (see \citet{DeMarco-2009} for a detailed review). This paper describes briefly  past and present efforts aimed at estimating an unbiased binary fraction of CSPNe.

\section{The  binary fraction obtained using photometric variability}

Periodic photometric variability of a binary CSPN is due to irradiation effect from the hot CSPN onto the companion, tidal deformations and eclipses \citep{Bond-2000,MiszalskiEtAl-2009}. Detecting photometric variability  requires repeated observations of targets  and can be done from the ground in non-photometric weather conditions. For this reason, it is an efficient binary-detection method and provides constantly new results. The main drawback of this method is that it is biased to small separations as irradiation effect, tidal deformations and eclipses all increase in intensity or frequency with decreasing separations. \\
\citet{Bond-2000} and \citet{MiszalskiEtAl-2009} already estimated close binary fractions  of CSPNe of respectively 10-15\% and 12-21\% with binary periods $\lesssim$~3 days. Although these fractions are lower limits for the overall binary fraction, their comparison with the  MS stars short-period binary fraction up to these these separations i.e. 5-7\% (\citealt{DuquennoyAndMayor-1991,RaghavanEtAl-2010,DeMarcoEtAl-2012}) already reveals that more PNe are formed around binaries. \\
Hillwig et al. (these proceedings) are monitoring targets from the 2.5 kpc volume-limited sample of \citet{Frew-2008} to estimate a new fraction of close binaries. Although their method is similar, the sample is less biased than the previously used magnitude-limited samples and also deeper (V $<$ 21). In a similar experiment, \citet{JacobyEtAl-2012} are monitoring  5-6 CSPNe within the Kepler satellite field of view to estimate an independent binary fraction. Their sample is statistically small ;  however the CSPNe are observed with a precision never reached before ($<$1 mmag).

\section{The binary fraction obtained  using red and infrared excess}
The red/IR excess technique aims to detect the signature of a cool, unresolved companion by measuring the photometry of the hot CSPN. To do so, high precision absolute photometry needing photometric weather conditions in the $B$, $V$ and $I$ (or $J$) bands is required. This technique is fully described in \citet{DeMarcoEtAl-2012}. The measured $B-V$ color is compared to the expected $B-V$ for the CSPN temperature according to atmospheric stellar models (e.g. \citealt{RauchAndDeetjen-2003}) and allows us to determine the reddening, whereas the $V-I$ or $V-J$ colours allows us to measure the red/IR excess, which is the difference between the $V-I$ or $V-J$ colour expected for a single star at the CSPN temperature and the  measured (de-reddened) one. If this difference is greater than the error on the photometric measurement, we list the object as a possible/probable binary. Since companions cooler than $\simeq$ M0-5 are faint, we need excellent photometric precision. Once a binary  fraction has been estimated, it can be compared to the MS one \citep{RaghavanEtAl-2010} only after undetected systems are accounted for. Using the $J$-band allows us to detect colder companions, while still not being contaminated by hot dust, although it requires a separate NIR observing run and is therefore time demanding.
\\\citet{FrewAndParker-2007} have used the photometry from the 2MASS and DENIS NIR  surveys to determine a  binary fraction $\simeq$~54\% but the detection bias was poorly quantified. \citet{DeMarcoEtAl-2012} have used the method described above on a sample of 27 CSPNe and have found a similar debiased fraction  $\simeq$~30\% from I-band data and $\simeq$~54\% from $J$-band data of a subset of 11 CSPNe in line with \citet{FrewAndParker-2007} result of 52-58\%. These fractions are biased in that companions fainter than M3-4V in I and M5-6V in J are not detected. Before this fraction can be compared with that of the mother population (the F6V-G2V stars, see \citet{RaghavanEtAl-2010} ; 50$\pm$4\%), a second bias must be taken into account ; by design the CSPN survey does not include resolved binaries which are instead accounted for in the MS binary fraction. Once these biases are accounted for, \citet{DeMarcoEtAl-2012} calculate a PN binary fraction of 70-100\%. The uncertainty is large because the statistics are poor. These  preliminary results will be confronted by the study of an additional 23 objects for which absolute photometry has been acquired at the NOAO 2.1m telescope in March 2011 as well as $\simeq$ 30 objects for which $J$ and $H$-band photometry has been obtained at the AAT 4m telescope in 2011 and the ANU 2.3m telescope in 2012. These new measurements should bring the sample to a statistically significant size and considerably reduce the error bars on the binary fraction. Recent surveys including $J$-band photometry will be analysed as well to extract the IR excess of other targets from the sample of \citet{Frew-2008}.

\section{The WD binary fraction and period distribution compared to those of the CSPN population}
The comparison of binary CSPN and binary WD populations can supply us with further constraints of the binary hypothesis. \citet{DeMarcoFarihiAndNordhaus-2009} ascertained that there are similarities between the period distributions of CSPN and WDs, although our knowledge of both is too coarse to draw detailed conclusions. In particular, since we presume that \emph{all} WDs come from the 1-8~M$_{\odot}$ population the WD binary fraction should truly reflect that of the main sequence, and be different from that of the CSPN population. The WD binary fraction of 30-40\% appears to be in line with that of the main sequence progenitor population (we recall that the progenitors of the entire WD sample have a median mass smaller than for a given CSPN and that the corresponding progenitor population binary fraction is therefore smaller as well, \citealt{RaghavanEtAl-2010}). It remains to be seen what the CSPN overall binary fraction is and whether it is indeed much larger than for the putative main sequence population and in line with the hypothesis that there is a preferential binary channel for PN formation.

\section{Conclusion}

Estimating an unbiased binary fraction of CSPNe is crucial to the understanding of whether companions play a key role in shaping PNe. Photometric variability has allowed us to determine a close binary fraction of $\simeq$ 15-20\% and is still being refined  on a new, less biased sample to understand  the biases inherent to this method. The red/IR excess technique leads us to obtain a CSPN binary fraction of 70-100\%, much larger than for the MS population. However, this number carries a large uncertainty for the moment due to the small sample size. Current studies based on optical and NIR photometry as well as the use of recent NIR surveys will double the sample size to constrain the CSPN binary fraction precisely enough to support or refute the hypothesis that PNe could emerge preferentially from  binary star evolution.

\bibliography{douchin-bibliography}

\end{document}